\documentclass{article}
\usepackage{spconf,amsmath,graphicx}

\usepackage{color}
\usepackage{enumitem}
\usepackage{hyperref}
\usepackage{multirow}
\usepackage{etoolbox,siunitx}
\robustify\bfseries
\usepackage{booktabs}
\usepackage{graphics}
\usepackage{balance}
\usepackage{amssymb}
\usepackage{bm}
\usepackage{cite}

\let\OLDthebibliography\thebibliography
\renewcommand\thebibliography[1]{
  \OLDthebibliography{#1}
  \setlength{\parskip}{2pt}
  \setlength{\itemsep}{3pt plus 0.5ex}
}

\title{Locate This, Not That: Class-Conditioned Sound Event DOA Estimation}
\name{Olga Slizovskaia, Gordon Wichern, Zhong-Qiu Wang, Jonathan Le Roux}
\address{Mitsubishi Electric Research Laboratories (MERL), Cambridge, MA, USA\\
{\small\texttt{\{wichern,leroux\}@merl.com}}}

\begin{document}
\maketitle
\begin{abstract}

Existing systems for sound event localization and detection (SELD) typically operate by estimating a source location for all classes at every time instant. In this paper, we propose an alternative class-conditioned SELD model for situations where we may not be interested in localizing all classes all of the time. This class-conditioned SELD model takes as input the spatial and spectral features from the sound file, and also a one-hot vector indicating the class we are currently interested in localizing. We inject the conditioning information at several points in our model using feature-wise linear modulation (FiLM) layers. Through experiments on the DCASE 2020 Task 3 dataset, we show that the proposed class-conditioned SELD model performs better in terms of common SELD metrics than the baseline model that locates all classes simultaneously, and also outperforms specialist models that are trained to locate only a single class of interest. We also evaluate performance on the DCASE 2021 Task 3 dataset, which includes directional interference (sound events from classes we are not interested in localizing) and notice especially strong improvement from the class-conditioned model.

\end{abstract}

\section{Introduction}
\label{sec:intro}

Estimating the direction of arrival (DOA) of a sound while also classifying the type of event is an important type of front-end processing for a wide variety of monitoring and robotic applications.  Due to its wide applicability, this task, often referred to in the literature as sound event localization and detection (SELD), has recently seen a surge of interest \cite{Adavanne2019, Politis2021, evers2020locata, grumiaux2021survey}.
However, SELD remains challenging because sound sources can move, cease to produce sound, have their positions obscured by room reverberation, and are often mixed with interfering sounds.  Furthermore, many sound events are easily confused, further complicating the SELD task.
   
A typical SELD pipeline consists of two main stages.  The first stage extracts spectral and spatial features from the microphone array input.  In the second stage, a deep neural network (DNN) is used to learn a mapping from the multi-channel input features to two output targets: (1) a vector of class probabilities indicating the presence of each class in each time instance; and (2) a DOA vector containing location information for each class.  In this case, DNN training can be challenging as the contributions from the two output branches must be balanced.  To overcome this difficulty, Shimada et al.~\cite{shimada2021accdoa} recently proposed the activity coupled Cartesian direction of arrival (ACCDOA) representation. ACCDOA represents the DOA information for each class as a vector of Cartesian coordinates, where the class activity probability is encoded by the length of the vector, i.e., it should be one when that class is active and zero otherwise.

\begin{figure}
    \centering
    \includegraphics[width=1\linewidth]{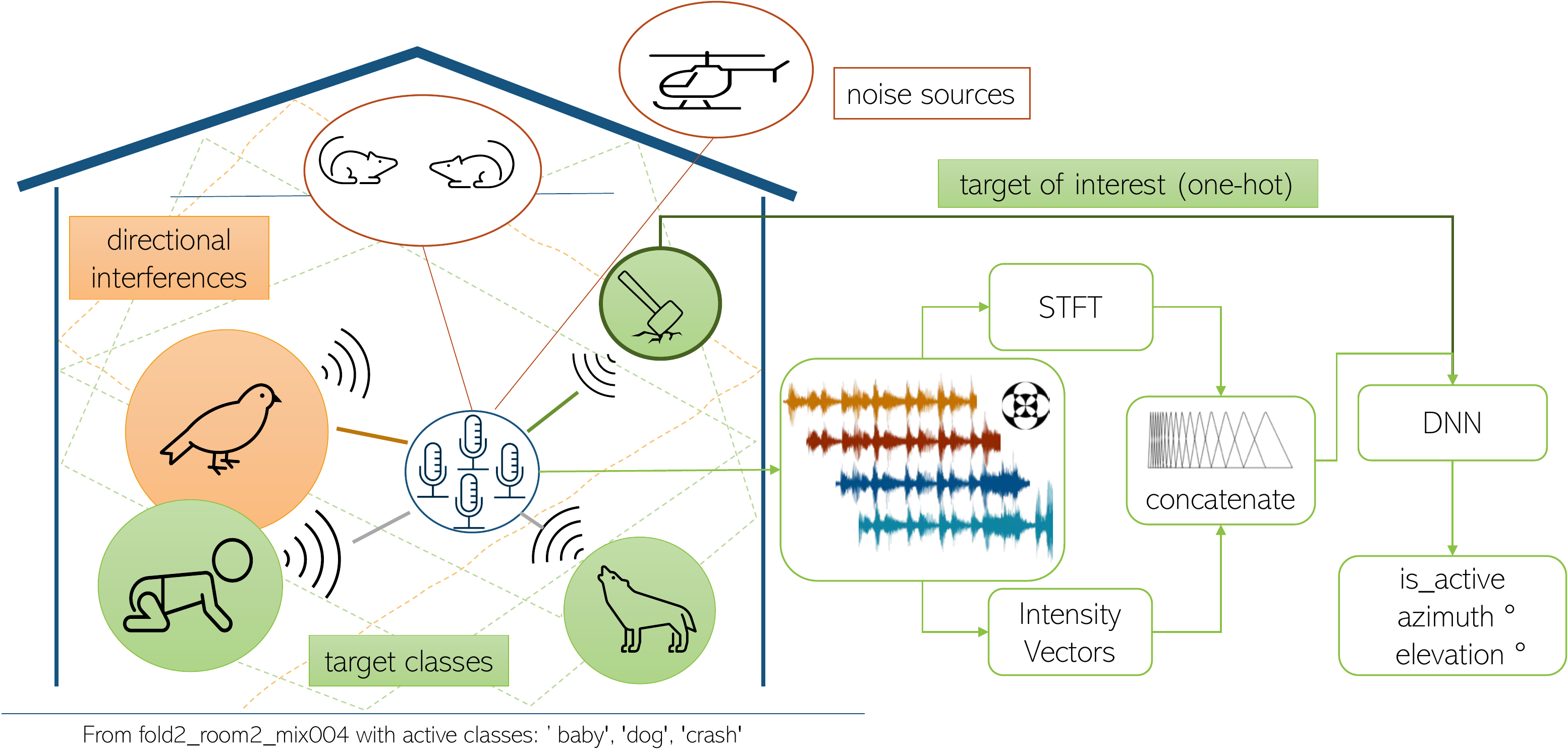}%
    \caption{Task illustration.
    Given a target-class identifier and a multi-channel signal recorded in conditions with multiple simultaneously active target sources, stationary noises, and directional interferences, we aim at estimating the DOA of the source of interest. Example taken from the training split of TAU-NIGENS SSE 2021 dataset.}%
    \label{fig:seld_problem}
\end{figure}

Existing SELD systems, including those based on the ACCDOA representation, typically assume a small, fixed set of sound event classes to be detected and localized.  In practice, if there are 12 classes in the vocabulary, this means a DNN will output 12 ACCDOA representations at each time instant.  This approach may become impractical for large class vocabularies, and, in certain situations, we may not be interested in localizing all classes all of the time.  
Training a class-specific system to localize only sound events from a single class would allow focusing on specific classes. However, this will then require $C$ class-specific models as opposed to a single one, if we are interested in localizing $C$ class types, and all models will also need to be run at inference time in order to detect all classes.  Furthermore, it may be difficult to train class-specific models, as we may not have enough data for each class of interest to properly train each model.

Inspired by the concepts of auditory attention~\cite{fritz2007auditory, huang2020push}, where the human auditory system tunes its internal models based on the sound sources it wants to listen to, in this paper we propose a class-conditioned approach for sound event DOA estimation, illustrated in Fig.~\ref{fig:seld_problem}. In our system, there is only a single ACCDOA vector output at each time instant, and the class represented by this output is determined based on an input describing the type of sound event we want to locate. Such conditioning-based approaches are often used in audio source separation~\cite{8683007} to isolate the sound from a specific musical instrument~\cite{Meseguer19CUNet, slizovskaia2021cunet}, speaker~\cite{delcroix2018single, wang2019voicefilter}, or sound event~\cite{ochiai2020listen}. Compared to a system that outputs all classes simultaneously, conditioning-based approaches require multiple passes at inference-time (one per class) in order to detect all classes, similarly to the class-specific models, although they do not require training multiple separate models. However, in situations where we may not want to detect all classes all of the time, they present several advantages.  First, they can easily scale up to large numbers of classes without having to add more parameters (i.e., additional output layers), and they also lend themselves to query-by-example~\cite{lee2019audio} or few-shot~\cite{wang2020few} settings where the conditioning input is not based on a class label embedding, but rather on an example of the type of sound we want to localize.

One other benefit of class-conditioned localization, which we demonstrate experimentally, is increased robustness to directional interference, i.e., sound events from classes we are not interested in localizing. We hypothesize that this increased robustness is a consequence of the class-conditioned training scheme, where all classes besides the one we are currently conditioning on are essentially treated as interfering sources, and this increased exposure allows the network to become better at ignoring directional interference. 
Through experiments on the DCASE 2020 Task 3 dataset~\cite{politis2020dataset}, which does not contain directional interference, we demonstrate that our proposed class-conditioned model performs better in terms of a set of common SELD metrics than a model that outputs all classes, and a set of class-specific models.  On the DCASE 2021 Task 3 dataset~\cite{politis2021dataset}, which does contain directional interference, the class-conditioned model outperforms more clearly the all-class model, while the class-specific models fail dramatically.

The rest of this paper is organized as follows. Section~\ref{sec:method} describes the baseline system, the class-specific SELD models, and our proposed class-conditioned SELD model.
Experimental setup and evaluation results are presented and discussed in Section~\ref{sec:experiments}, followed by conclusions in Section~\ref{sec:conclusion}.

\section{Method}
\label{sec:method}
We first review the baseline SELD architecture with ACCDOA output representation, and then describe the proposed class-conditioned modification.

\subsection{Baseline: All-class and Class-specific SELD models}
The ACCDOA representation~\cite{shimada2021accdoa} uses Cartesian coordinates for the class-dependent azimuth and elevation angles, which are the estimation targets. Formally, let $\mathbf{d}_{t,c}=[x_{t,c}, y_{t,c}, z_{t,c}]$ represent the DOA for sound event class $c$ at frame $t$. If no sound events from class $c$ are active at frame $t$, $\|\mathbf{d}_{t,c}\|_2=0$, whereas if class $c$ is active, $||\mathbf{d}_{t,c}||_2=1$. In the all-class baseline model with $C$ classes of interest for a $T$-frame sound file, the tensor $\mathbf{D}\in\mathbb{R}^{T \times C \times 3}$ represents all ACCDOA vectors, and the goal is to estimate DNN parameters $\theta$ to estimate $\mathbf{D}$, i.e., $\hat{\mathbf{D}}=\mathcal{F^{\mathrm{ac}}}_{\theta}(\mathbf{X})$, where $\mathbf{X}\in\mathbb{R}^{M \times F \times T}$ represents the $M$-microphone, $F$-dimensional input features. Parameters $\theta$ are estimated by minimizing the mean squared error (MSE) between $\mathbf{D}$ and $\hat{\mathbf{D}}$. At inference time, class $c$ is declared active at time $t$ if $||\hat{\mathbf{d}}_{t,c}||_2 > \tau$, where $\tau$ is a threshold typically chosen to optimize performance on the validation set.

For the class-specific SELD models, the goal is to learn, for each class $c$, model parameters $\theta_{c}$ in order to obtain DOA estimates $\hat{\mathbf{D}}_c\!=\!\mathcal{F^{\mathrm{cs}}}_{\theta_{c}}(\mathbf{X})\!\in\!\mathbb{R}^{T \times 3}$ for class $c$. Each $\theta_{c}$ is learned by minimizing the MSE between $\mathbf{D}_c\!=\!\mathbf{D}[:,c,:]$ and $\hat{\mathbf{D}}_c$. At inference time, we use the same threshold for all models. %

\subsection{Class-conditioned SELD model}
\label{ssec:cond_seld}

One potential weakness of the baseline all-class SELD scheme is a high sparsity of the target representation tensor $\mathbf{D}$ and the need to predict all DOAs at once.
In this case, we ask the network to not only focus on all locations but also identify all classes of interest, solving two different but interconnected problems.
Instead, to solve the simpler task of predicting a DOA of a particular target class, it may be beneficial to focus the network on that class through an auxiliary conditioning input. %
Formally, we first define a $C$-dimensional one-hot vector $\mathbf{o}_c=[o_1,\dots,o_C]$, where $o_c=1$ for the class $c$ we are interested in localizing, and $o_i=0, \forall i \neq c$. Our goal is now to learn model parameters $\theta$ in order to estimate $\hat{\mathbf{D}}_c=\mathcal{F^{\mathrm{cc}}}_{\theta}(\mathbf{X}, \mathbf{o}_c)$, where $\hat{\mathbf{D}}_c \in\mathbb{R}^{T \times 3}$ represents the DOA estimates for class $c$ only. Contrary to the class-specific models which each have their own parameter space, the class-conditioned model shares the same set of parameters for all classes. During training, we concatenate the outputs from evaluating $\mathcal{F^{\mathrm{cc}}}_{\theta}(\mathbf{X}, \mathbf{o}_c)$ for all $C$ classes, such that we can obtain the full $\hat{\mathbf{D}}\in\mathbb{R}^{T \times C \times 3}$, and as in the all-class system we can learn $\theta$ by minimizing the mean squared error between $\mathbf{D}$ and $\hat{\mathbf{D}}$.  Note that we have here $C$ training examples for each multi-channel signal instead of just one, i.e., a training example is now a pair $(\mathbf{X},\mathbf{o}_c)$. 

\begin{figure}
    \centering
    \includegraphics[width=\linewidth]{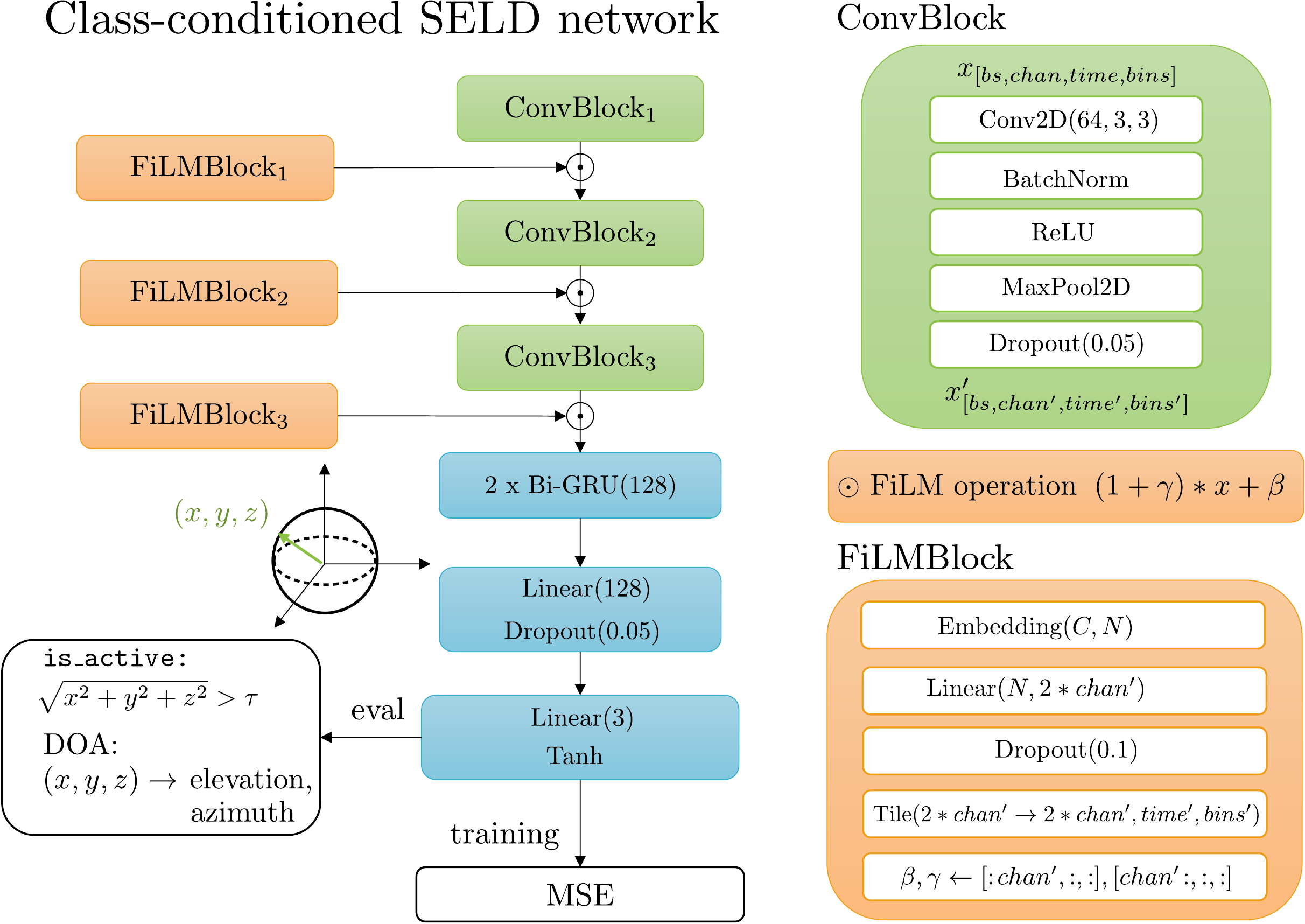}
    \caption{Illustration of class-conditioned SELD architecture. Feature maps $x'_{[bs, chan^{'}, time^{'}, bins^{'}]}$ of each convolutional block of the baseline SELD model are shifted and scaled using a set of parameters $(\gamma, \beta)$ learned from a conditioning one-hot vector.}
    \label{fig:filmed_seld}
\end{figure}

Conditioning mechanisms, in particular feature-wise transformations, have proven to be a simple and effective technique for providing a model with context information. For our class-conditioned SELD model, we employ the feature-wise linear modulation (FiLM)~\cite{dumoulin2018featurewise} scheme. We first pass the one-hot vector $\mathbf{o}_c$ through a learned embedding layer, to obtain a class embedding $\mathbf{e}\in\mathbb{R}^N$ that encodes the target class information. 
A FiLM layer $f$ outputs a set of parameters $ (\gamma, \beta) = f(\mathbf{e})$ that are used for scaling and shifting the learned feature maps $\mathbf{M}$ of the baseline network: $ \text{FiLM}(\mathbf{M}_i|\gamma_{i}, \beta_{i}) = \gamma_{i}\mathbf{M}_i + \beta_{i} $, where $i$ refers to a channel index. As shown in Fig.~\ref{fig:filmed_seld}, we apply the FiLM operation to the feature maps of each convolutional block in the baseline SELD network. While a FiLM transformation can consist of any arbitrary function, in practice, we use a combination of an embedding, linear, and dropout layers as shown in Fig.~\ref{fig:filmed_seld}.

\section{Experimental validation}
\label{sec:experiments}

\subsection{Experimental Setup}
We evaluate our method on two official development datasets for 
the DCASE SELD task: TAU-NIGENS Spatial Sound Events 2020 \cite{politis2020dataset} and TAU-NIGENS Spatial Sound Events 2021 \cite{politis2021dataset}. For both datasets, we follow the official split into training (400 minutes), validation (100 minutes), and test (100 minutes) sets. The two key differences between the datasets are a reduced number of classes and the presence of directional interferences in the 2021 edition. The TAU-NIGENS 2021 dataset consists of mixtures of sound events belonging to 12 different classes, while the TAU-NIGENS 2020 dataset includes two more target classes,  namely ``engine'' and ``fire.'' As reported in \cite{politis2021datasetpaper}, the 2021 edition has more simultaneously active events (up to three parallel tracks of different or same categories), and additional directional interferences from non-target classes. In all experiments, we use a recommended set of features extracted from the first-order ambisonic (FOA) data split that includes 4-channel 64-dimensional mel-spectrograms and 3-channel FOA intensity vectors. 

Our network architecture follows the convolutional recurrent architecture used as the DCASE SELD baseline~\cite{politis2021datasetpaper} and shown in Fig.~\ref{fig:filmed_seld}, but the FiLM layers are only included in the class-conditioned models.  Both the class-conditioned and class-specific models have only a single ACCDOA output, while the all-class baseline has $C$ ACCDOA outputs. We train all models for 50 epochs (note that an epoch is $C$ times larger for the class-conditioned model, because each class and sound file pair counts as a data sample as discussed in Section~\ref{ssec:cond_seld}) using the ADAM optimizer with a learning rate of $1e^{-3}$.  We note that our goal for the experiments in this section is not to explore the data augmentation~\cite{wang2021four} and model ensembles~\cite{shimada2021ensemble} necessary to obtain the best scores on the DCASE SELD datasets, but rather explore the impact of class-conditioned SELD models in a controlled manner. 

Following the DCASE SELD setup, we use four metrics and an aggregated score to evaluate different aspects of each SELD system. The four metrics are (1) the location-sensitive error rate $ER_{20^{\circ}}$ that only accepts the predicted event as true positive if its location is at most $20^{\circ}$ away from the ground truth and accounts for substitutions, deletions, and insertions from other classes, (2) the location-sensitive $F$-measure $F_{20^{\circ}}$, (3) the class-sensitive localization error $LE_{CD} $ that is computed separately for each class, and (4) the class-sensitive localization recall $LR_{CD} $. The aggregated SELD score~\cite{mesaros2019seldmetrics} provides a simplified metric for the overall evaluation and is computed as follows:
\begin{equation}
SELD =  \frac{ER_{20^{\circ}}}{4} + \frac{1 - F_{20^{\circ}}}{4} + \frac{LE_{CD}}{4\times180} + \frac{1 - LR_{CD}}{4}.
\end{equation}

The ACCDOA format uses a norm of the DOA vector $\|\mathbf{d}_{t,c}\|_2$ to decide if class $c$ is active at frame $t$ or not. The default threshold $\tau$ on the norm is set to $0.5$. However, the optimal choice of the threshold is a practical issue that is raised in \cite{shimada2021ensemble}. Our empirical evaluation shows that the difference in SELD scores can be as large as $0.04$ points depending on the threshold. For a fair comparison, we select an optimal threshold in terms of the SELD score for each experiment based on the validation set, and report the best results that a system can achieve together with the chosen threshold.

\subsection{Results}

\begin{table*}[ht!]
    \centering
  \sisetup{table-format=2.2,round-mode=places,round-precision=2,table-number-alignment = center,detect-weight=true,detect-inline-weight=math}
      \caption{Experimental results for baselines and the proposed system evaluated on the testing fold of the development datasets for DCASE 2021 Task 3 and DCASE 2020 Task 3.}\vspace{.05cm}
\resizebox{\linewidth}{!}
{%
    \begin{tabular}{l
    S[table-format=1.1,round-precision=1]S[table-format=1.2]SSSS
    S[table-format=1.1,round-precision=1]S[table-format=1.2]SSSS}
    \toprule
         & \multicolumn{6}{c}{DCASE 2020} & \multicolumn{6}{c}{DCASE 2021} \\
         \cmidrule(lr){2-7} \cmidrule(lr){8-13} 
         System & {$\tau$} & {$ER_{20^{\circ}}\!\downarrow $} & {$F_{20^{\circ}}\!\uparrow $} & {$LE_{CD}\!\downarrow $} & {$LR_{CD}\!\uparrow $} & {$SELD\!\downarrow $ } & {$\tau$} & {$ER_{20^{\circ}}\!\downarrow $} & {$F_{20^{\circ}}\!\uparrow $} & {$LE_{CD}\!\downarrow $} & {$LR_{CD}\!\uparrow $} & {$SELD\!\downarrow $ } \\ 
         \midrule
         Baseline FOA~\cite{politis2021dataset} & 0.4 & 0.75 & 37.5 & 23.3 & 53.0 & 0.494 & 0.3 & 0.71 & 36.24 & 23.20 & 47.34 & 0.501 \\
         Specialized models & 0.5 & 0.79 & 35.1 & 22.3 & 46.9 & 0.52 & 0.5 & 2.22 & 21.7 & 23.2 & 54.4 & 0.90 \\
         Class-conditioned SELD & 0.3 & 0.72 & 47.9 & 17.8 & 55.9 & 0.445 & 0.3 & 0.64 & 49.72 & 19.16 & 56.60 & 0.420 \\
         \bottomrule
    \end{tabular}
    }
\label{tab:results}
\vspace{-.5cm}
\end{table*}

\begin{figure}
    \centering
    \includegraphics[width=1\linewidth]{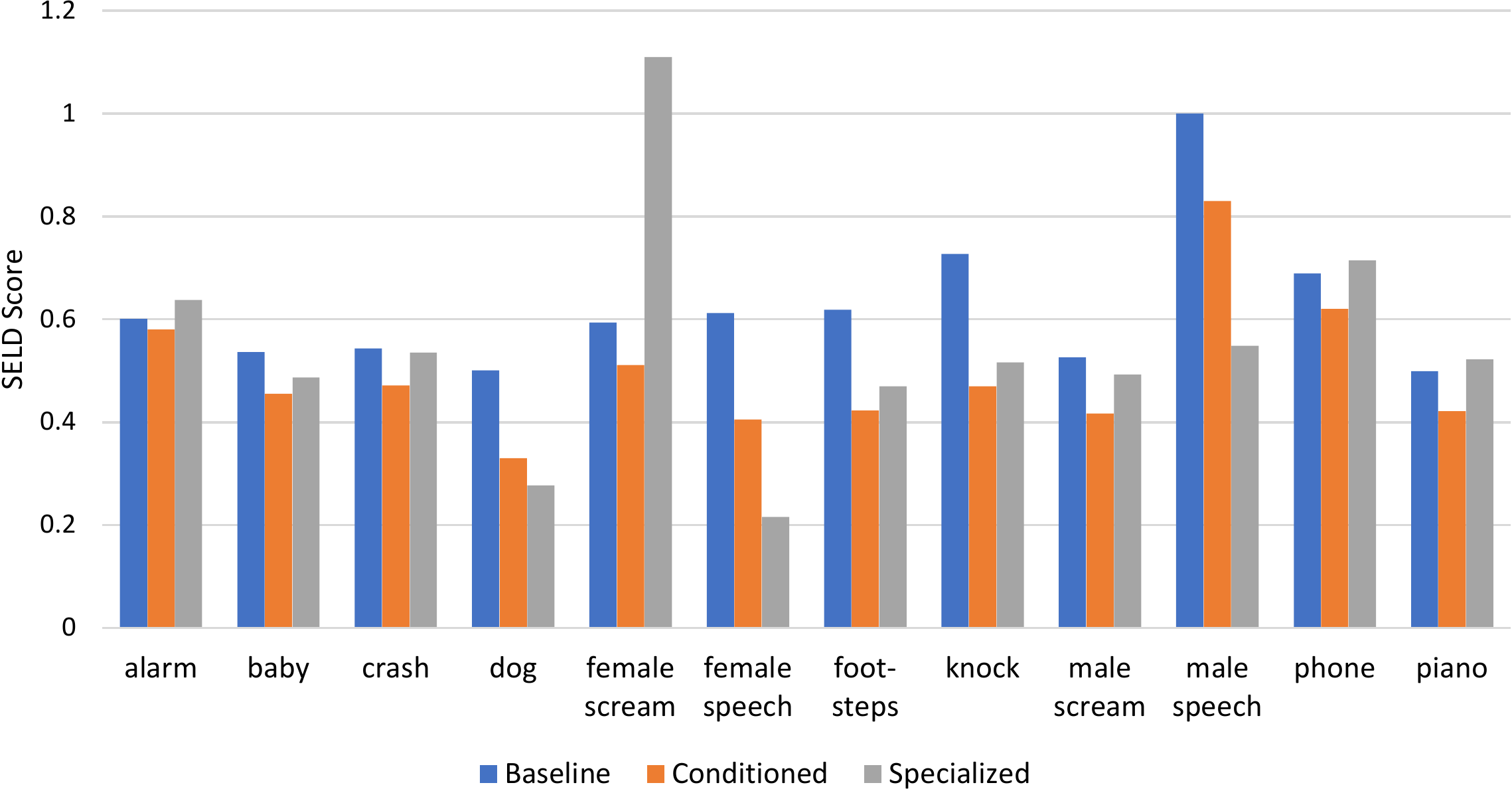}
    \caption{Comparison of per-class performance for different models on TAU-NIGENS 2021 data.}
    \label{fig:per_class_performance}
\vspace{-.5cm}
\end{figure}

Results for the baseline all-class SELD network, the class-specific models, and the proposed class-conditioned SELD model are presented in Table~\ref{tab:results}.
As the original baseline is allegedly not fully reproducible, we ran it six times with different random seeds for the DCASE 2021 dataset to make sure there was not too much variance in the results, and report averaged performance. As we found that the measured standard deviation on the DCASE 2021 data was in fact low, we only report the results from a single run for the DCASE 2020 dataset. 
For the class-specific models, to obtain the aggregated scores shown in Table~\ref{tab:results}, we collect predictions from the model for each class and compare them against the ground truth. Similarly, we collect predictions of the proposed class-conditioned SELD model ran with class-conditionings for each class. 

When comparing performance between the two baselines and the proposed class-conditioned SELD model, we observe that the class-conditioned model outperforms both baselines in terms of the SELD score and all individual metrics on both the DCASE 2021 Task 3 and DCASE 2020 Task 3 datasets. The relative reduction in terms of the SELD score is 16\% for the DCASE 2021 dataset and 8.2\% for the DCASE 2020 dataset compared to the baseline. We attribute a higher gain obtained for the DCASE 2021 data to the fact that the class-conditioned model is more robust to directional interferences which are present in the TAU-NIGENS 2021 dataset but not present in the TAU-NIGENS 2020 dataset.

\begin{figure}
    \centering
    \includegraphics[width=1\linewidth]{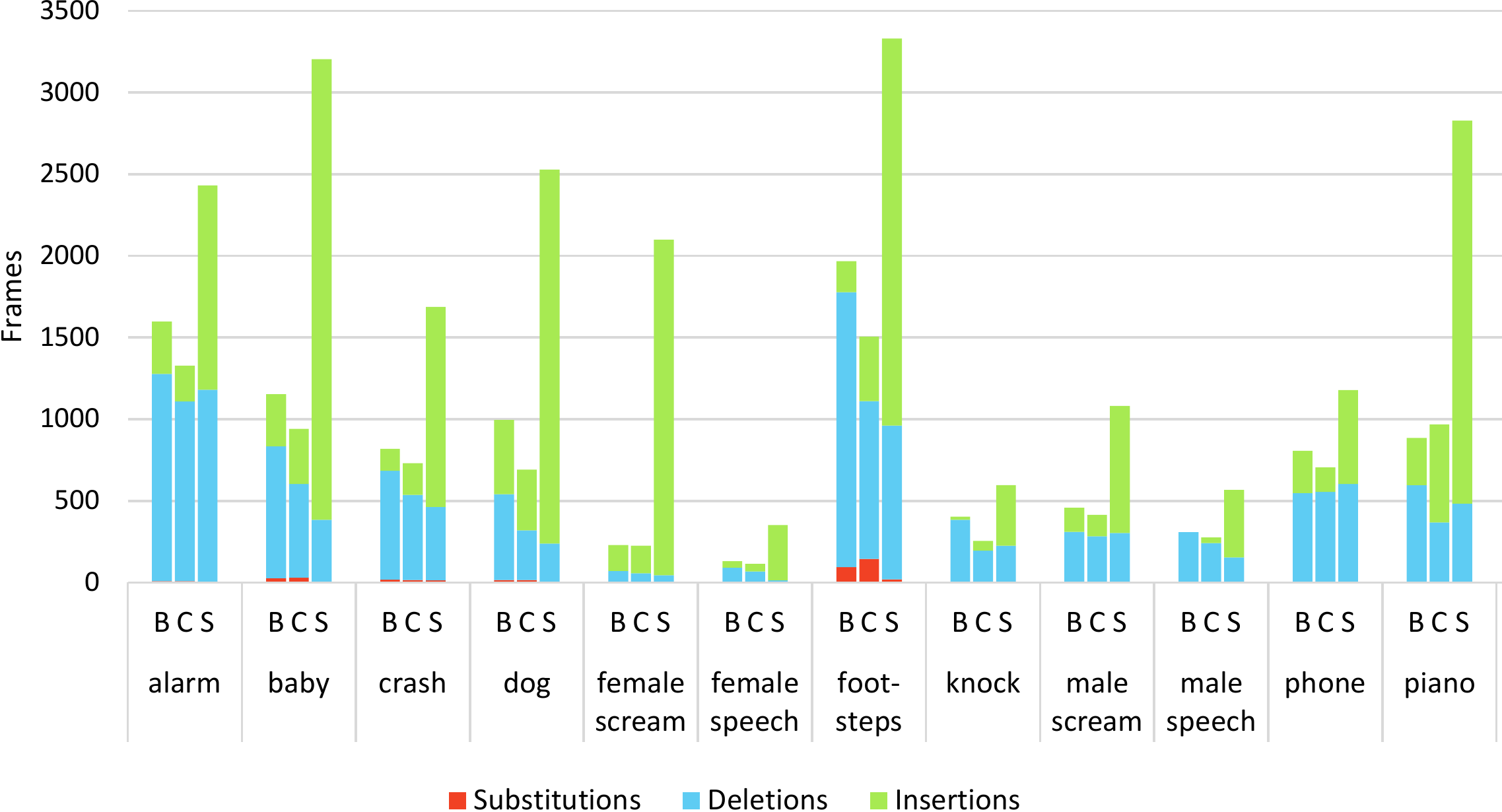}
    \caption{Per-class comparison of the number of frames estimated as substitutions, deletions, and insertions for the baseline (B), class-conditioned (C) and specialized (S) systems on the DCASE 2021 Task 3 dataset. }
    \label{fig:per_class_sdi}
\vspace{-.3cm}
\end{figure}

Figure~\ref{fig:per_class_performance} illustrates per-class performance in terms of the SELD score for different systems on the DCASE 2021 Task 3 dataset. Interestingly, despite the fact that the aggregated SELD metrics for the specialized models demonstrate worse results, we still observe that class-specific systems can yield better scores than both the baseline and class-conditioned models for ``dog,'' ``female speech,'' and ``male speech'' categories. To further investigate this issue, we consider individual substitution, deletion, and insertion scores for each system and class. Figure~\ref{fig:per_class_sdi} shows that the specialized models have a significantly higher number of insertions compared to the baseline and class-conditioned systems. While the effect of insertions on individual per-class SELD scores is less noticeable in Fig.~\ref{fig:per_class_performance} (with the only distinct exception being the ``female scream'' class), the accumulated effect of all insertions makes the location-sensitive error rate three times worse compared to the baseline. We also note that the specialized model performance in Table~\ref{tab:results} for the DCASE 2020 dataset does not exhibit the same large drop observed for the DCASE 2021 dataset. Since the DCASE 2021 dataset contains more overlapping sources,  training a model that only uses labels for a single-class may be difficult if that class is rarely observed in isolation.

\vspace{-.3cm}
\section{Conclusion}
\label{sec:conclusion}
\vspace{-.2cm}

In this study, we focus on sound event detection and localization in the presence of directional interferences. We propose a class-conditioned SELD network that outperforms the baseline on the DCASE 2020 Task 3 and DCASE 2021 Task 3 datasets. The proposed system demonstrates greater improvement for the DCASE 2021 Task 3 dataset, supporting our hypothesis that a conditioned network is more robust against directional interferences such as those included in this dataset. The class-conditioned SELD network can be easily extended to accommodate a larger number of target categories which is one of the possible directions for future research. 
Other possible extensions of this work include incorporation of data augmentation techniques as proposed in~\cite{wang2021four}, zero-shot sound event localization with pre-trained class-specific embeddings, as well as integration of a source separation module as proposed in~\cite{masahiro2020localizationbyseparation} or a source counting module.

\bibliographystyle{IEEEtran}
\bibliography{refs}

% Generated by IEEEtran.bst, version: 1.13 (2008/09/30)
\begin{thebibliography}{10}
\providecommand{\url}[1]{#1}
\csname url@samestyle\endcsname
\providecommand{\newblock}{\relax}
\providecommand{\bibinfo}[2]{#2}
\providecommand{\BIBentrySTDinterwordspacing}{\spaceskip=0pt\relax}
\providecommand{\BIBentryALTinterwordstretchfactor}{4}
\providecommand{\BIBentryALTinterwordspacing}{\spaceskip=\fontdimen2\font plus
\BIBentryALTinterwordstretchfactor\fontdimen3\font minus
  \fontdimen4\font\relax}
\providecommand{\BIBforeignlanguage}[2]{{%
\expandafter\ifx\csname l@#1\endcsname\relax
\typeout{** WARNING: IEEEtran.bst: No hyphenation pattern has been}%
\typeout{** loaded for the language `#1'. Using the pattern for}%
\typeout{** the default language instead.}%
\else
\language=\csname l@#1\endcsname
\fi
#2}}
\providecommand{\BIBdecl}{\relax}
\BIBdecl

\bibitem{Adavanne2019}
S.~Adavanne, A.~Politis, J.~Nikunen, and T.~Virtanen, ``Sound event
  localization and detection of overlapping sources using convolutional
  recurrent neural networks,'' \emph{IEEE Journal on Selected Topics in Signal
  Processing}, vol.~13, no.~1, pp. 34--48, 2019.

\bibitem{Politis2021}
A.~Politis, A.~Mesaros, S.~Adavanne, T.~Heittola, and T.~Virtanen, ``Overview
  and evaluation of sound event localization and detection in {DCASE} 2019,''
  \emph{IEEE/ACM Trans. Audio, Speech, Lang. Process.}, vol.~29, pp. 684--698,
  2021.

\bibitem{evers2020locata}
C.~Evers, H.~W. L{\"o}llmann, H.~Mellmann, A.~Schmidt, H.~Barfuss, P.~A.
  Naylor, and W.~Kellermann, ``The {LOCATA} challenge: Acoustic source
  localization and tracking,'' \emph{IEEE/ACM Trans. Audio, Speech, Lang.
  Process.}, vol.~28, pp. 1620--1643, 2020.

\bibitem{grumiaux2021survey}
P.-A. Grumiaux, S.~Kiti{\'{c}}, L.~Girin, and A.~Gu{\'{e}}rin, ``A survey of
  sound source localization with deep learning methods,'' \emph{arXiv preprint
  arXiv:2109.03465}, 2021.

\bibitem{shimada2021accdoa}
K.~Shimada, Y.~Koyama, N.~Takahashi, S.~Takahashi, and Y.~Mitsufuji,
  ``{ACCDOA}: Activity-coupled {C}artesian direction of arrival representation
  for sound event localization and detection,'' in \emph{Proc. ICASSP}, Jun.
  2021, pp. 915--919.

\bibitem{fritz2007auditory}
J.~B. Fritz, M.~Elhilali, S.~V. David, and S.~A. Shamma, ``Auditory
  attention—focusing the searchlight on sound,'' \emph{Current opinion in
  neurobiology}, vol.~17, no.~4, pp. 437--455, 2007.

\bibitem{huang2020push}
N.~Huang and M.~Elhilali, ``Push-pull competition between bottom-up and
  top-down auditory attention to natural soundscapes,'' \emph{ELife}, vol.~9,
  p. e52984, 2020.

\bibitem{8683007}
P.~Seetharaman, G.~Wichern, S.~Venkataramani, and J.~Le~Roux,
  ``Class-conditional embeddings for music source separation,'' in \emph{Proc.
  ICASSP}, May 2019, pp. 301--305.

\bibitem{Meseguer19CUNet}
G.~Meseguer-Brocal and G.~Peeters, ``Conditioned-{U-Net}: Introducing a control
  mechanism in the {U-Net} for multiple source separations,'' in \emph{Proc.
  ISMIR}, Nov. 2019.

\bibitem{slizovskaia2021cunet}
O.~Slizovskaia, G.~Haro, and E.~Gomez~Gutierrez, ``Conditioned source
  separation for musical instrument performances,'' \emph{IEEE/ACM Trans.
  Audio, Speech, Lang. Process.}, vol.~29, pp. 2083--2095, 2021.

\bibitem{delcroix2018single}
M.~Delcroix, K.~Zmolikova, K.~Kinoshita, A.~Ogawa, and T.~Nakatani, ``Single
  channel target speaker extraction and recognition with speaker beam,'' in
  \emph{Proc. ICASSP}, Apr. 2018, pp. 5554--5558.

\bibitem{wang2019voicefilter}
Q.~Wang, H.~Muckenhirn, K.~Wilson, P.~Sridhar, Z.~Wu, J.~Hershey, R.~A.
  Saurous, R.~J. Weiss, Y.~Jia, and I.~L. Moreno, ``{VoiceFilter}: Targeted
  voice separation by speaker-conditioned spectrogram masking,'' in \emph{Proc.
  Interspeech}, Sep. 2019.

\bibitem{ochiai2020listen}
T.~Ochiai, M.~Delcroix, Y.~Koizumi, H.~Ito, K.~Kinoshita, and S.~Araki,
  ``Listen to what you want: Neural network-based universal sound selector,''
  in \emph{Proc. Interspeech}, Oct. 2020.

\bibitem{lee2019audio}
J.~H. Lee, H.-S. Choi, and K.~Lee, ``Audio query-based music source
  separation,'' in \emph{Proc. ISMIR}, Nov. 2019.

\bibitem{wang2020few}
Y.~Wang, J.~Salamon, N.~J. Bryan, and J.~P. Bello, ``Few-shot sound event
  detection,'' in \emph{Proc. ICASSP}, May 2020, pp. 81--85.

\bibitem{politis2020dataset}
\BIBentryALTinterwordspacing
A.~Politis, S.~Adavanne, and T.~Virtanen, ``{TAU-NIGENS Spatial Sound Events
  2020},'' May 2020. [Online]. Available:
  \url{https://doi.org/10.5281/zenodo.4064792}
\BIBentrySTDinterwordspacing

\bibitem{politis2021dataset}
\BIBentryALTinterwordspacing
------, ``{TAU-NIGENS Spatial Sound Events 2021},'' Feb. 2021. [Online].
  Available: \url{https://doi.org/10.5281/zenodo.4844825}
\BIBentrySTDinterwordspacing

\bibitem{dumoulin2018featurewise}
V.~Dumoulin, E.~Perez, N.~Schucher, F.~Strub, H.~de~Vries, A.~Courville, and
  Y.~Bengio, ``Feature-wise transformations,'' \emph{Distill}, 2018,
  https://distill.pub/2018/feature-wise-transformations.

\bibitem{politis2021datasetpaper}
A.~Politis, S.~Adavanne, D.~Krause, A.~Deleforge, P.~Srivastava, and
  T.~Virtanen, ``A dataset of dynamic reverberant sound scenes with directional
  interferers for sound event localization and detection,'' \emph{arXiv
  preprint arXiv:2106.06999}, 2021.

\bibitem{wang2021four}
Q.~Wang, J.~Du, H.-X. Wu, J.~Pan, F.~Ma, and C.-H. Lee, ``A four-stage data
  augmentation approach to {ResNet}-{Conformer} based acoustic modeling for
  sound event localization and detection,'' \emph{arXiv preprint
  arXiv:2101.02919}, 2021.

\bibitem{shimada2021ensemble}
``Ensemble of {ACCDOA- and EINV2-based} systems with {D3Nets} and impulse
  response simulation for sound event localization and detection.''

\bibitem{mesaros2019seldmetrics}
A.~Mesaros, S.~Adavanne, A.~Politis, T.~Heittola, and T.~Virtanen, ``Joint
  measurement of localization and detection of sound events,'' in \emph{Proc.
  WASPAA}, Oct. 2019, pp. 333--337.

\bibitem{masahiro2020localizationbyseparation}
M.~Yasuda, Y.~Koizumi, S.~Saito, H.~Uematsu, and K.~Imoto, ``Sound event
  localization based on sound intensity vector refined by {DNN}-based denoising
  and source separation,'' in \emph{Proc. ICASSP}, May 2020, pp. 651--655.

\end{thebibliography}

\end{document}